\documentclass{jps-cp}
\usepackage{xspace}
\usepackage{amsmath}
\usepackage[T1]{fontenc}
\newcommand{\pp}           {pp\xspace}

\newcommand{\PbPb}         {\mbox{Pb--Pb}\xspace}

\newcommand{\pPb}          {\mbox{p--Pb}\xspace}

\newcommand{\s}            {\ensuremath{\sqrt{s}}\xspace}

\newcommand{\pt}           {\ensuremath{p_{\rm T}}\xspace}

\newcommand{\etarange}[1]  {\mbox{$\left | \eta \right |<#1$}}

\newcommand{\dndeta}       {\ensuremath{\mathrm{d}N_\mathrm{ch}/\mathrm{d}\eta}\xspace}
\newcommand{\avdndeta}     {\ensuremath{\langle\dndeta\rangle}\xspace}

\newcommand{\pns}          {\ensuremath{{\rm P}(n_{\rm S})}\xspace}
\newcommand{\inelgz}       {\ensuremath{{\rm INEL>0}}\xspace}

\newcommand{\acceff}       {\ensuremath{A\kern-.15em\times\kern-.15em\varepsilon}\xspace}

\newcommand{\nineH}        {$\sqrt{s}~=~0.9$~Te\kern-.1emV\xspace}
\newcommand{\seven}        {$\sqrt{s}~=~7$~Te\kern-.1emV\xspace}
\newcommand{\twoH}         {$\sqrt{s}~=~0.2$~Te\kern-.1emV\xspace}
\newcommand{\twosevensix}  {$\sqrt{s}~=~2.76$~Te\kern-.1emV\xspace}
\newcommand{\five}         {$\sqrt{s}~=~5.02$~Te\kern-.1emV\xspace}
\newcommand{\twosevensixnn}{$\sqrt{s_{\mathrm{NN}}}~=~2.76$~Te\kern-.1emV\xspace}
\newcommand{\fivenn}       {$\sqrt{s_{\mathrm{NN}}}~=~5.02$~Te\kern-.1emV\xspace}

\newcommand{\GeVc}         {Ge\kern-.1emV/$c$\xspace}
\newcommand{\MeVc}         {Me\kern-.1emV/$c$\xspace}
\newcommand{\TeV}          {Te\kern-.1emV\xspace}
\newcommand{\GeV}          {Ge\kern-.1emV\xspace}
\newcommand{\MeV}          {Me\kern-.1emV\xspace}
\newcommand{\GeVmass}      {Ge\kern-.2emV/$c^2$\xspace}
\newcommand{\MeVmass}      {Me\kern-.2emV/$c^2$\xspace}

\newcommand{\kzero}        {\ensuremath{{\rm K}^{0}_{\rm{S}}}\xspace}
\newcommand{\lmb}          {\ensuremath{\Lambda}\xspace}
\newcommand{\almb}         {\ensuremath{\overline{\Lambda}}\xspace}
\newcommand{\Om}           {\ensuremath{\Omega^-}\xspace}
\newcommand{\Mo}           {\ensuremath{\overline{\Omega}^+}\xspace}
\newcommand{\X}            {\ensuremath{\Xi^-}\xspace}
\newcommand{\Ix}           {\ensuremath{\overline{\Xi}^+}\xspace}

\newcommand{\XiNosign}           {\ensuremath{\Xi}\xspace}
\newcommand{\OmNosign}           {\ensuremath{\Omega}\xspace}

\newcommand{\Xzero}            {\ensuremath{\Xi^0}\xspace}

\newcommand{\dndetaNew}       {\ensuremath{{\langle\dndeta\rangle}_{\etarange{0.5}}}~(V0M~class)\xspace}

\newcommand{\ynp}[2]  {\ensuremath{\rm \langle Y_{#1#2} \rangle}\xspace}
\newcommand{\dels}       {\ensuremath{{\rm \Delta S}}\xspace}

\newcommand{\pem}       {PYTHIA~8 Monash 2013\xspace}
\newcommand{\pecrr}     {PYTHIA~8 QCD-CR + Ropes\xspace}
\newcommand{\eplhc}     {EPOS~LHC\xspace}

\title{Probing Strange-Quark Hadronization via (Multi-)Strange Hadron Multiplicity Distributions in Small Collision Systems with ALICE}

\author{Sara \textsc{Pucillo}$^{1}$ for the ALICE Collaboration}

\inst{$^{1}$ Dipartimento di Fisica, Università degli Studi di Salerno, and Sezione INFN Salerno, Via Giovanni Paolo II, 132, 84084 Fisciano SA, Italy}

\email{sara.pucillo@cern.ch}

\begin{document}

\abst{Strangeness enhancement is defined as the increased relative production of strange hadrons in heavy-ion collisions compared to proton--proton (pp) interactions. It was originally proposed as one of the signatures of quark--gluon plasma (QGP) formation. At the LHC, the ALICE experiment observed that strange-hadron-to-pion yield ratios rise with increasing charged-particle multiplicity at midrapidity, independently of collision energy (\s) and system size, from pp to \pPb and up to \PbPb collisions. To gain deeper insight into the mechanisms of strangeness production, the ALICE collaboration has measured the probability distribution of producing a given number of strange particles (\kzero, \lmb (\almb), \X (\Ix), and \Om (\Mo)) of the same species per event in \pp collisions at \five~\cite{ALICE:2025strange}. This measurement extends the study of strangeness production beyond the mean particle yield by employing, for the first time, a technique based on event-by-event particle counting. It provides a new test bench for production mechanisms, probing events with large imbalances between strange and non-strange content. The results are compared with state-of-the-art phenomenological models implemented in commonly used Monte Carlo event generators, offering enhanced sensitivity to the underlying dynamics of strangeness production.}

\kword{Strangeness, ALICE, small systems}

\maketitle

\section{Introduction} \label{intro}

Strange-hadron production provides a sensitive tool to investigate hadronization in high-energy collisions, since it probes how strange quarks are incorporated into final-state hadrons during the non-perturbative stage of QCD. In heavy-ion collisions, the relative enhancement of strange and multi-strange hadrons with respect to non-strange particles has long been regarded as one of the characteristic observations associated with the formation of a deconfined medium~\cite{rafelski}. This behavior was first established at the SPS~\cite{Andersen:1999,Afanasiev:2002,Antinori:2010} and later confirmed at RHIC~\cite{starSE} and at the LHC~\cite{PbPb2,PbPb5}.

A particularly intriguing development of the LHC program is that signatures traditionally discussed in the context of heavy-ion collisions have also emerged in high-multiplicity small systems. In this framework, ALICE measurements of strange-hadron-to-pion yield ratios have shown that strange-particle production evolves smoothly with charged-particle multiplicity across \pp, \pPb, and \PbPb collisions~\cite{nature,pp5,pp7,pp13,pPbV0,pPbCasc,PbPb2,PbPb5}. This result indicates that event activity plays a central role in shaping strange-particle production, and it has stimulated renewed interest in the mechanisms governing hadron formation in dense final states. The observation of strangeness enhancement in \pp collisions, together with other collective-like effects reported in high-multiplicity \pp and \pPb interactions~\cite{ridge1,ridge2,ridgecms,flowcms1,flowcms2,flowatlas,alicecollaboration2024}, has therefore broadened the scope of strangeness studies beyond the traditional heavy-ion picture.

On the theory side, these findings have motivated significant developments in Monte Carlo event generators, where mechanisms such as color reconnection, rope hadronization, or modified cluster fragmentation have been introduced to account for the observed behavior in high-density environments~\cite{pythiainfo,herwigmanual,PythiaCRbaryon,pythiaRopes,herwigstrangeness}. At the same time, several dedicated measurements have explored how strange-particle production is connected to other event properties, such as jet activity~\cite{lk0sjets,2024}, event isotropy~\cite{ALICEspherocity}, effective energy~\cite{2025}, and strange quantum-number conservation~\cite{2024BalanceF,PoissMario2025}. 

\section{Measurement of (multi-)strange particle multiplicity distribution} \label{Pns}

To better understand how strangeness is produced, the ALICE collaboration has measured the probability distribution for producing a given number of strange particles, \pns, of the same species in a single event in \pp collisions at \five, using a new technique based on event-by-event strange-particle counting~\cite{ALICE:2025strange}.

After topological selections on the weak decay and particle identification of the decay daughters, signal and combinatorial background contributions are separated through fits to the invariant-mass distributions in intervals of \pt and event multiplicity. The signal shape is described by a double-sided Crystal Ball (dsCB) function~\cite{Gaiser:1982yw}, while the residual background is parameterized with a first-degree polynomial (Gaussian for the \OmNosign). This allows one to assign to each candidate a signal probability weight,
\begin{equation}
    w_{\it S}~(\pt{\rm ,{\it m}_{inv};V0M}) = \frac{{\rm dsCB}~(\pt{\rm ,{\it m}_{inv};V0M})}{{\rm Total}~(\pt{\rm ,{\it m}_{inv};V0M})}~\{~\cdot~w_{FD}(\pt{\rm ;V0M})~\}~~ \quad , ~~ w_{\it B}~=~1-w_{\rm S},
    \label{eq}
\end{equation}
with an additional feed-down correction factor $w_{FD}$ applied in the \lmb(\almb) case to account for contributions from \X (\Ix) and \Xzero decays.

Each of the $N$ candidates in a given event is considered together with its corresponding signal weight $w_{\it S}$. The probabilities of all possible signal--background configurations are then evaluated and summed, from the case in which all candidates are background to the one in which all are signal. In this way, an event-by-event probability distribution for the true strange-particle multiplicity, ranging from 0 to $N$, is reconstructed. The final \pns distribution is obtained by summing these event-level probabilities over the full data sample and normalizing to the total number of analyzed events.

Detector effects are corrected through a one-dimensional Bayesian unfolding procedure~\cite{unf}, based on dedicated simulations including realistic strange-particle \pt spectra and detector conditions. In addition, trigger inefficiency is accounted for both at the event level, through the factor $\varepsilon_{trig}$ reported in Ref.~\cite{multEtrig5}, and at the particle level, through a correction factor $\varepsilon_{part}$ estimated from MC simulations for candidates lost in non-triggered events. The corrected multiplicity distribution is therefore written as
\begin{equation}
 {\rm P}(n_{\rm S}) = \frac{\varepsilon_{trig}}{\varepsilon_{part}^{n}} \cdot \left\{ \frac{\sum_{i=1}^{N_{tot}}R_{i}[n]}{N_{tot}}\right\}_{Unfolded},
 \label{eq_yields}
\end{equation}
where $R_{i}[n]$ is the reconstructed distribution and $N_{tot}$ is the total number of analyzed events.

\begin{figure}[tbh]
  \centering
  \includegraphics[width=0.48\linewidth]{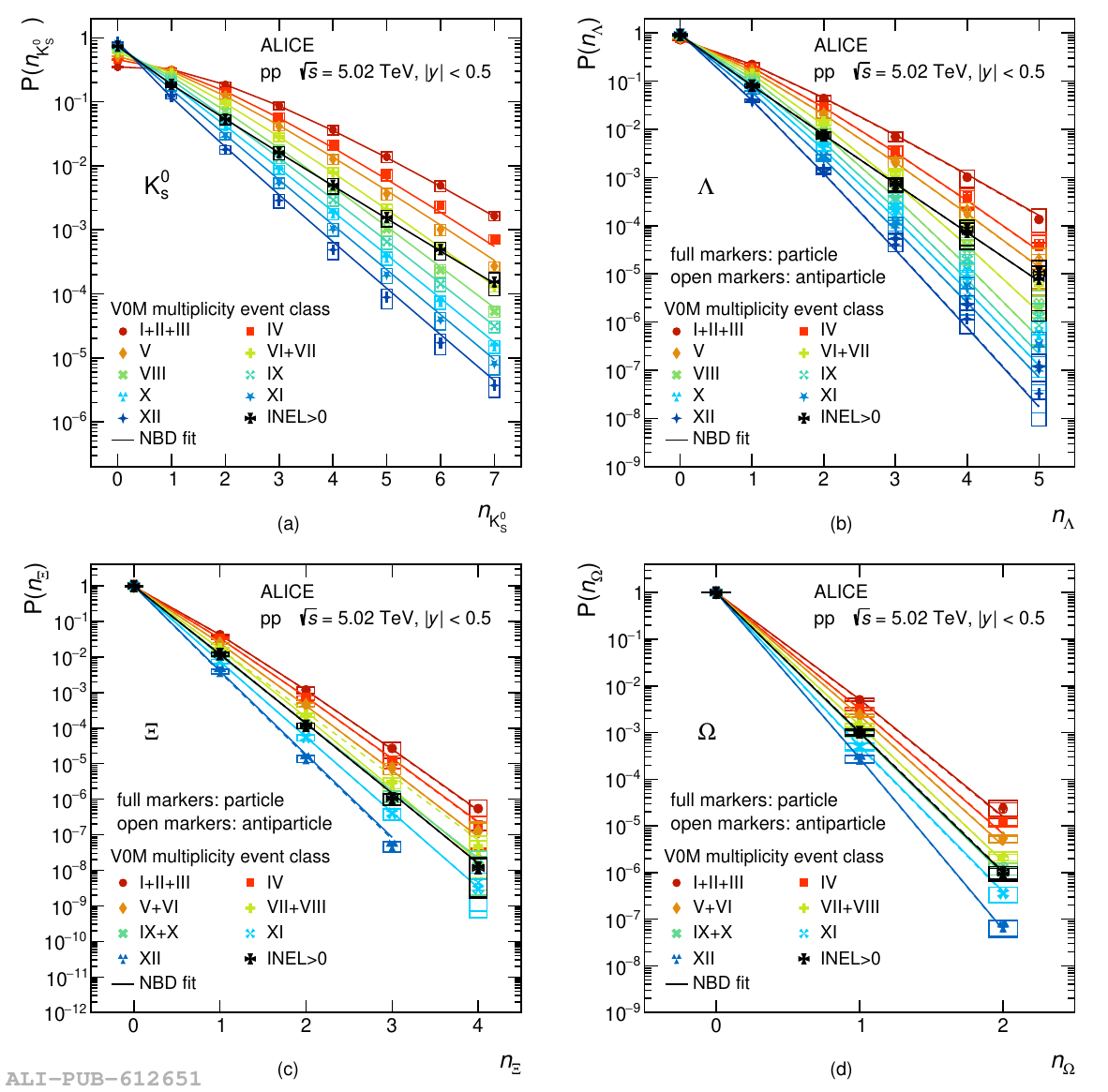}
  \hfill
  \includegraphics[width=0.48\linewidth]{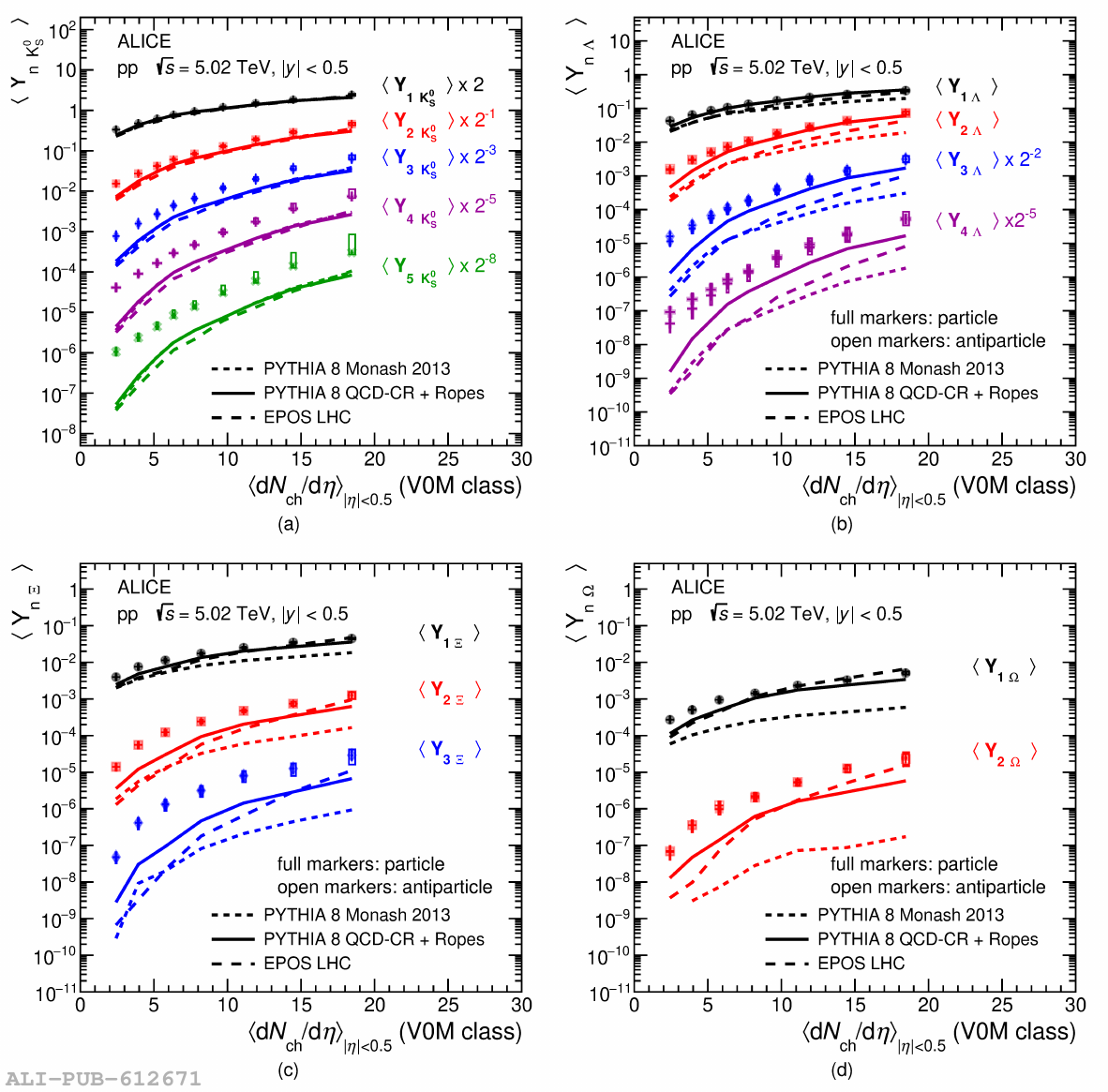}
  \caption{(\textit{left}): (Multi-)strange-particle multiplicity distributions, \pns, for \kzero (a), \lmb and \almb (b), \X and \Ix (c), and \Om and \Mo (d) in several multiplicity classes. Continuous (dashed) lines show the NBD fit~\cite{NBDfit} for particles (antiparticles). (\textit{right}): Multiple strange-hadron production yields as a function of the average charged-particle multiplicity \dndetaNew for \kzero (a), \lmb (b), \XiNosign (c), and \OmNosign (d). Full markers indicate particles, open markers antiparticles. Predictions from \pem, \pecrr, and \eplhc are shown as dotted, continuous, and dashed lines, respectively.}
  \label{pnsyield}
\end{figure}

The resulting \pns distributions are shown in Fig.~\ref{pnsyield} (\textit{left}) for several multiplicity classes, including \inelgz, i.e. events with at least one reconstructed silicon-pixel tracklet in \etarange{1}, corresponding to about 75\% of the total inelastic cross section. The available data sample allows the measurement of \pns up to 7 \kzero, 5 \lmb, 4 \XiNosign (3 in the lowest multiplicity interval), and 2 \OmNosign particles per event. This provides a unique opportunity to study the connection between charged and strange particle production in extreme configurations, ranging from events with 7 \kzero at low average multiplicity, where \avdndeta $\sim 3$, to events with zero \kzero at high multiplicity, where \avdndeta $\sim 20$.

For all species, the probability of observing $n>0$ strange hadrons increases with multiplicity, consistently with the previously established rise of average strange-particle yields~\cite{pp5}. The separation between low- and high-multiplicity classes becomes larger at increasing $n$, highlighting a stronger-than-linear growth in the high-multiplicity tail. The same distributions are also well described by a Negative Binomial Distribution (NBD) fit~\cite{NBDfit}, in analogy with earlier charged-particle multiplicity measurements~\cite{Pnch09,ValentinaPnch}, with $\chi^2$ values below 1 for V0s and below 3 for cascades.

\section{Multiple strange hadron production yields and yield ratios} \label{yields}

From the measurement of \pns, the average production yield of $n$ strange particles per event (multiplets of order $n$) can be calculated as
\begin{equation}
  \ynp{n}{S} = \sum_{i=n}^{n_{S}^{max}} \frac{i!}{n!(i-n)!} \cdot P(i_{S}),
  \label{eq:yields}
\end{equation}
where $n_{\mathrm{S}}^{max}$ is the highest measured bin of the multiplicity distribution for particle $S$. The combinatorial factor accounts for all combinations of $i_S$ strange particles contributing to the order-$n$ yield:
\begin{align*}
  &\ynp{1}{S} = P(1_{S}) + 2\cdot P(2_S) + 3\cdot P(3_S) + \cdots\\
  &\ynp{2}{S} = P(2_S) + 3\cdot P(3_S) + 6\cdot P(4_S) + \cdots\\
  &\ynp{3}{S} = P(3_S) + 4\cdot P(4_S) + 10\cdot P(5_S) + \cdots\\
  &\cdots
\end{align*}

The corresponding yields are shown in Fig.~\ref{pnsyield} (\textit{right}) for \kzero, \lmb, \XiNosign, and \OmNosign as a function of charged-particle multiplicity. \ynp{1}{S} corresponds to the average of the multiplicity distribution, while \ynp{n>1}{S} identifies the average production yield of multiplets of the same particle. The increase with multiplicity is more than linear for multiple strange-hadron production. Comparisons with Pythia 8 Monash~\cite{Pythia}, Pythia 8 (QCD-CR) Ropes~\cite{PythiaRopes}, and Epos LHC~\cite{EPOS} show that the description worsens as the number of strange particles per event increases.

\begin{figure}[tbh]
  \centering
  \includegraphics[width=0.45\linewidth]{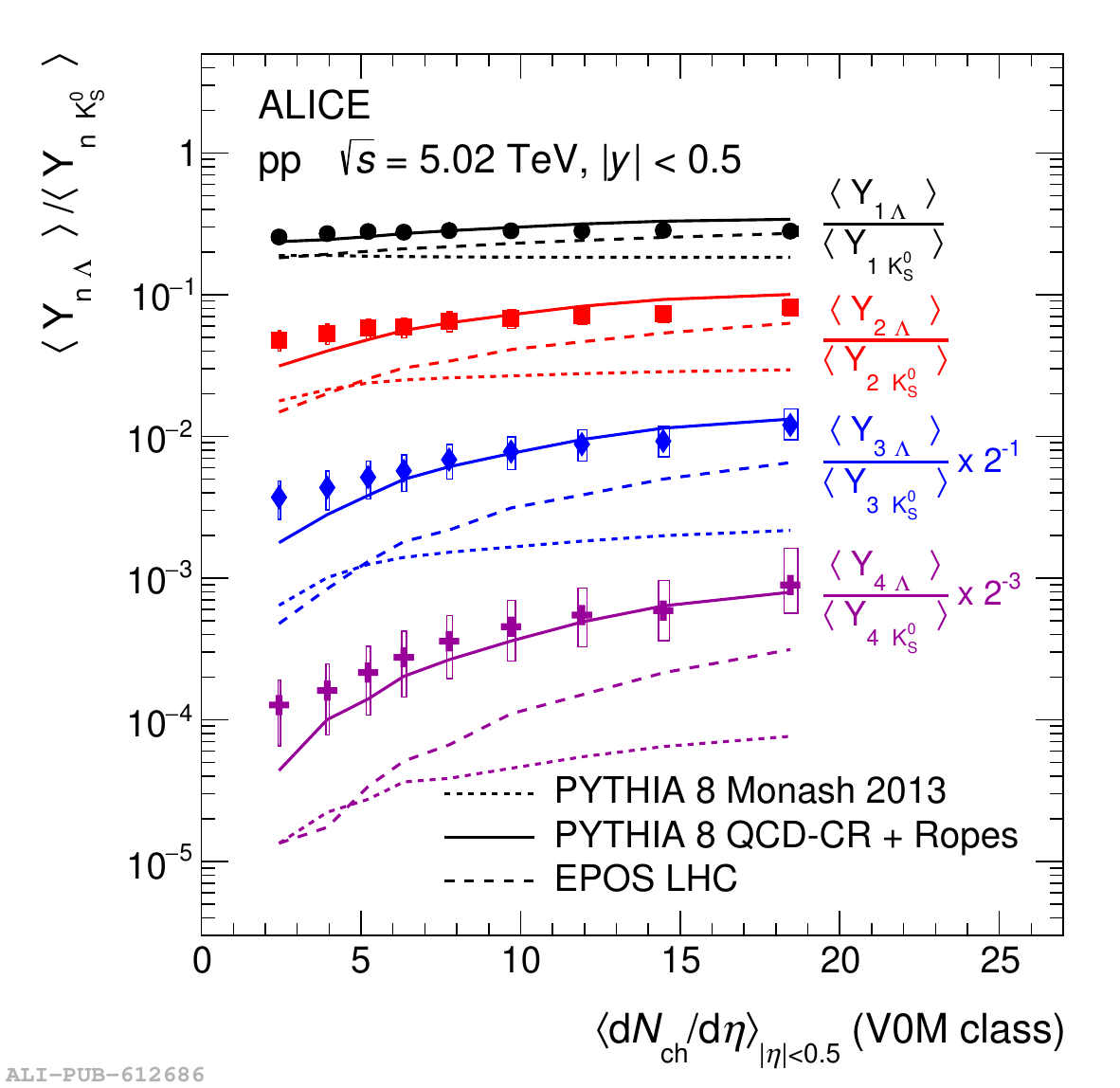}

  \vspace{0.3cm}

  \includegraphics[width=0.9\linewidth]{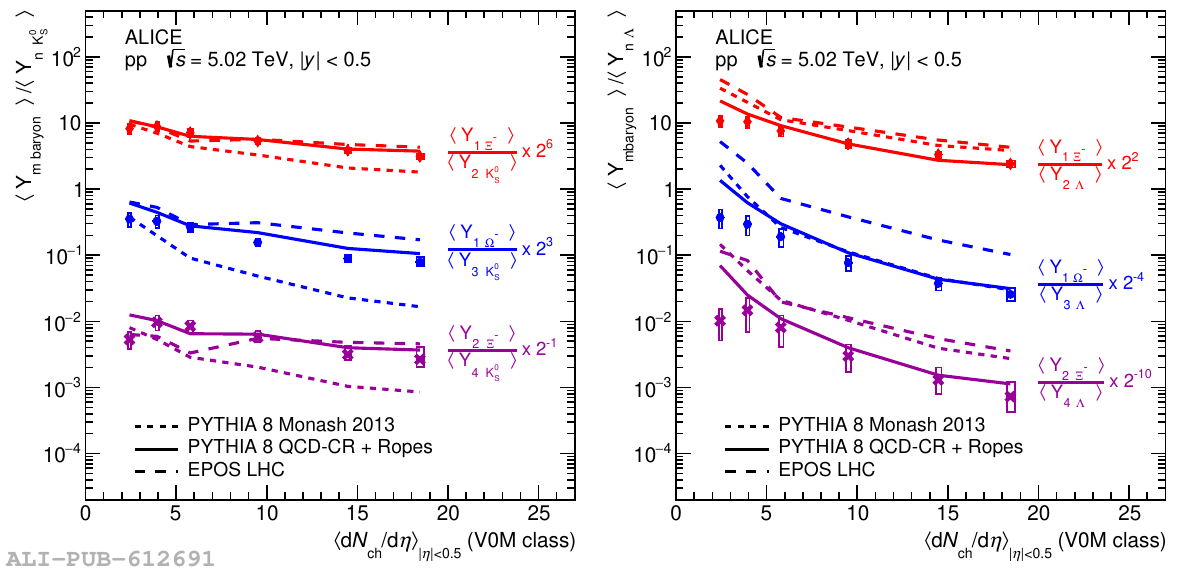}
  \caption{(\textit{top}): \ynp{n}{\lmb}/\ynp{n}{\kzero} as a function of the average charged-particle multiplicity at midrapidity, for $n=1$ to 4 from top to bottom. (\textit{bottom}): Ratios of the average production yield of $m$ multi-strange baryons to those of $n$ single-strange particles, using \kzero (\textit{left}) or \lmb (\textit{right}) in the denominator, as a function of \dndetaNew. The combinations shown are \ynp{1}{\X}/\ynp{2}{\kzero}, \ynp{1}{\Om}/\ynp{3}{\kzero}, and \ynp{2}{\X}/\ynp{4}{\kzero} (\textit{left}), and \ynp{1}{\X}/\ynp{2}{\lmb}, \ynp{1}{\Om}/\ynp{3}{\lmb}, and \ynp{2}{\X}/\ynp{4}{\lmb} (\textit{right}), from top to bottom. Predictions from \pem, \pecrr, and \eplhc are shown as dotted, continuous, and dashed lines, respectively.}
  \label{yieldratios}
\end{figure}

The measurement of multiple strange-hadron yields makes it possible to investigate particle-yield ratios with either balanced or strongly unbalanced strange-quark content between numerator and denominator. Ratios with \dels = 0 are especially useful to isolate effects that are not trivially driven by net strangeness content. 
As shown in Fig.~\ref{yieldratios}, even perfectly balanced ratios (\dels = 0) exhibit a non-trivial multiplicity dependence. In particular, \ynp{n}{\lmb}/\ynp{n}{\kzero} increases with \dndetaNew for $n>1$, demonstrating for the first time that the evolution of multi-particle yield ratios with multiplicity cannot be attributed solely to the strangeness difference between numerator and denominator. 
Additional \dels = 0 ratios, shown in the second line of Fig.~\ref{yieldratios}, display the opposite behavior, but help clarify the picture. In particular, \ynp{1}{\X}/\ynp{2}{\kzero}, \ynp{1}{\Om}/\ynp{3}{\kzero}, and \ynp{2}{\X}/\ynp{4}{\kzero} decrease with multiplicity despite the heavier and more baryon-rich numerator. This suggests that neither mass nor baryon number alone determines the observed evolution. A qualitatively consistent interpretation is provided by a na\"ive quark-coalescence picture. At fixed strange-quark content, these ratios probe how the available strange quarks compete with the available light quarks in forming hadrons with different valence-quark structure. For example, forming an \Om requires only strange quarks, while forming three \kzero requires, in addition, three light quarks. In low-multiplicity events, the scarcity of light quarks increases the likelihood that the three strange quarks will form an \Om rather than three \kzero. Conversely, at higher multiplicity, the increasing abundance of light quarks makes kaon production more likely, leading to a suppression of ratios such as \ynp{1}{\Om}/\ynp{3}{\kzero}. A similar argument also accounts for the stronger decrease observed when baryons appear in the denominator, as in the ratios to \lmb, since a \lmb requires two light quarks whereas a \kzero requires only one.

Overall, these measurements show that the multiplicity evolution of multi-strange yield ratios at \dels = 0 contains non-trivial information beyond conventional strangeness enhancement and that the interplay between strange- and light-quark availability may provide a coherent qualitative explanation of the observed patterns. Model comparisons provide further insight into the underlying production mechanisms. \pem does not reproduce either the absolute values or the multiplicity evolution of the measured ratios. \eplhc gives a reasonable description for ratios to \kzero, but systematically overestimates those to \lmb. \pecrr provides the best overall agreement, reproducing both the absolute scale of the observables and their evolution. This suggests that, while the production rate of strange quarks remains an open issue, as indicated by the failure to describe the strange-hadron multiplet yields in Fig.~\ref{pnsyield} (right), the color-reconnection mechanism with light quarks captures the main trends observed in the data once strangeness is produced.

\section{Conclusions} \label{conclusions}

In summary, the production of (multi-)strange hadrons in \pp collisions at \five has been investigated using a novel event-by-event counting technique, extending the study of strangeness beyond average yields~\cite{ALICE:2025strange}. The measurement of \pns and the subsequent extraction of multiplet yields \ynp{n}{S} provide a new, stringent test for hadronization models, probing events with large imbalances between strange and non-strange content.
The results presented in this work, focusing on yield ratios with balanced strangeness content (\dels = 0), reveal a non-trivial multiplicity dependence that cannot be attributed to the strangeness content alone. The observed increase of \ynp{n}{\lmb}/\ynp{n}{\kzero} with multiplicity for $n > 1$, contrasted with the decreasing trends in ratios involving multi-strange baryons (such as \ynp{1}{\Om}/\ynp{3}{\kzero} and \ynp{m}{baryon}/\ynp{n}{\lmb}), suggests that light-quark availability plays a significant role in the hadronization process. These patterns are qualitatively consistent with a quark-coalescence picture, where the competition between strange and light quarks for hadron formation evolves with event activity. Among the tested models, \pecrr fornisce the most accurate description of these \dels = 0 trends, highlighting the importance of color reconnection mechanisms. 
Comprehensive results reported in Ref.~\cite{ALICE:2025strange}, which include unbalanced \dels ratios, allow for the study of strangeness enhancement at its extremes, reaching \dels up to 5 and an enhancement of two orders of magnitude from low to high multiplicity. Taken together, these measurements establish a new experimental approach that significantly constrains the dynamics of strange-particle production in small collision systems.

\end{document}